\UseRawInputEncoding
\documentclass[preprint]{aastex631}

\graphicspath{{./}{figures/}}

\begin{document}

\title{Clues to the formation of Liller 1 from modeling its complex star formation history.}

\author{Emanuele Dalessandro}
\affiliation{Istituto Nazionale di Astrofisica (INAF), Osservatorio di Astrofisica e Scienza dello Spazio di Bologna, Via Gobetti 93/3, Bologna
I-40129, Italy}
\author{Chiara Crociati}
\affiliation{Dipartimento di Fisica e Astronomia, Universit\'a  di Bologna, Via Gobetti 93/2, Bologna I-40129, Italy}
\affiliation{Istituto Nazionale di Astrofisica (INAF), Osservatorio di Astrofisica e Scienza dello Spazio di Bologna, Via Gobetti 93/3, Bologna
I-40129, Italy}
\author{Michele Cignoni}
\affiliation{Department of Physics, University of Pisa, Largo Pontecorvo, 3, I-56127 Pisa, Italy}
\affiliation{INFN, Largo B. Pontecorvo 3, I-56127 Pisa, Italy}
\affiliation{Istituto Nazionale di Astrofisica (INAF), Osservatorio di Astrofisica e Scienza dello Spazio di Bologna, Via Gobetti 93/3, Bologna
I-40129, Italy}
\author{Francesco R. Ferraro}
\affiliation{Dipartimento di Fisica e Astronomia, Universit\'a  di Bologna, Via Gobetti 93/2, Bologna I-40129, Italy}
\affiliation{Istituto Nazionale di Astrofisica (INAF), Osservatorio di Astrofisica e Scienza dello Spazio di Bologna, Via Gobetti 93/3, Bologna
I-40129, Italy}
\author{Barbara Lanzoni}
\affiliation{Dipartimento di Fisica e Astronomia, Universit\'a  di Bologna, Via Gobetti 93/2, Bologna I-40129, Italy}
\affiliation{Istituto Nazionale di Astrofisica (INAF), Osservatorio di Astrofisica e Scienza dello Spazio di Bologna, Via Gobetti 93/3, Bologna
I-40129, Italy}
\author{Livia Origlia}
\affiliation{Istituto Nazionale di Astrofisica (INAF), Osservatorio di Astrofisica e Scienza dello Spazio di Bologna, Via Gobetti 93/3, Bologna
I-40129, Italy}
\author{Cristina Pallanca}
\affiliation{Dipartimento di Fisica e Astronomia, Universit\'a  di Bologna, Via Gobetti 93/2, Bologna I-40129, Italy}
\affiliation{Istituto Nazionale di Astrofisica (INAF), Osservatorio di Astrofisica e Scienza dello Spazio di Bologna, Via Gobetti 93/3, Bologna
I-40129, Italy}
\author{R. Michael Rich}
\affiliation{Department of Physics and Astronomy, University of California, Los Angeles, Los Angeles, CA, USA}
\author{Sara Saracino}
\affiliation{Astrophysics Research Institute, Liverpool John Moores University, Liverpool, UK}
\author{Elena Valenti} 
\affiliation{European Southern Observatory, Garching, Germany}
\affiliation{Excellence Cluster ORIGINS, Garching, Germany}

\begin{abstract}
Liller 1 and Terzan 5 are two massive systems in the Milky-Way bulge hosting populations characterized by 
significantly different ages ($\Delta t>7-8$ Gyr) and metallicities ($\Delta$[Fe/H]$\sim1$ dex). 
Their origin is still strongly debated in the literature and all formation scenarios proposed 
so far require some level of fine-tuning. The detailed star formation histories (SFHs) of these systems 
may represent an important piece of information to assess their origin. Here we present the first attempt 
to perform such an analysis for Liller 1. The first key result we find is that Liller 1 has been forming 
stars over its entire lifetime. More specifically, three broad SF episodes are clearly detected: 
1) a dominant one, occurred some 12-13 Gyr ago with a tail extending for up to $\sim3$ Gyr, 
2) an intermediate burst, between 6 and 9 Gyr ago, 3) and a recent one, occurred between 1 and 3 Gyr ago. 
The old population contributes to about $70\%$ of the total stellar mass and the remaining 
fraction is almost equally split between the intermediate and young populations. 
If we take these results at a face value, 
they would suggest that this system unlikely formed through the merger between an old globular 
cluster and a Giant Molecular Cloud, as recently proposed. On the contrary, our findings 
provide further support to the idea that Liller 1 is the surviving relic of a 
massive primordial structure that contributed to 
the Galactic bulge formation, similarly to the giant clumps observed in star-forming high-redshift galaxies.

\end{abstract}

\keywords{Globular star clusters, Galactic bulge, Star formation, Photometry}

\section{Introduction}
Among the vast population of Galactic stellar aggregates traditionally classified as globular clusters, three systems, namely Terzan~5, Liller~1 and $\omega$ Centauri, 
stand out for their significantly peculiar properties. 
With masses larger than some $10^6 M_{\odot}$, these objects are among the most massive cluster-like systems in the Galaxy. 
More importantly, 
they are the only ones hosting multiple populations 
characterized by both significant iron and age spreads ($\sim1$ dex and up to $5-8$ Gyr respectively - e.g., \citealt{norris96,lee99,pancino00,ferraro09,ferraro21}).

$\omega$ Centauri is a metal-poor (average metallicity $<[Fe/H]>\sim-1.5$) system orbiting the Galactic halo 
and hosting at least 4-5 discrete populations, as revealed by a large number of 
detailed photometric and spectroscopic analyses 
(e.g., \citealt{norris96,lee99,pancino00,ferraro04,sollima05,bellini09,johnson10}). 
These populations span a wide metallicity range, going from [Fe/H]$\sim-2.2$ dex to $-0.5$ dex, and a possible 
age range of about $3-4$ Gyr. 
This evidence led to the early suggestion that $\omega$ Centauri is the remnant core of an accreted 
dwarf galaxy captured during its approach to the Milky Way \citep{zinnecker88,majewski00}.
Indeed, more recently, thanks to the exquisite Gaia astrometry \citep{brown21}, $\omega$ Centauri has been 
associated with either Gaia-Enceladus or 
Sequoia \citep{massari19}, 
which are two of the most prominent accretion events experienced by the Galaxy. 

On the contrary, both Terzan~5 and Liller~1 are suspected to be the result of a different and very peculiar 
formation process. Terzan~5, for which detailed high-resolution photometric and spectroscopic analyses have been performed 
\citep{ferraro09,origlia11,massari14}, hosts at least three sub-populations: {\it i)} a dominant old population 
that formed about 12 Gyr ago from gas enriched by type II supernovae (SNe), with sub-solar metallicity ([Fe/H]$= -0.3$) 
and enhanced [$\alpha$/Fe] abundance ratio; {\it ii)} a younger population formed about 4.5 Gyr ago, 
from gas characterized by super-solar metallicity ([Fe/H]$=+0.3$) and solar-scaled [$\alpha$/Fe]; {\it iii)}
an additional metal poor [Fe/H]$=-0.7$ and enhanced [$\alpha$/Fe] abundance ratio.
The [$\alpha$/Fe]-[Fe/H] pattern of Terzan~5 is strikingly similar to that of bulge stars \citep{ness13,johnson13,
gonzalez15,rojas17,queiroz20}, and totally 
incompatible with what observed in the Milky Way outer disk and halo, and in local dwarf galaxies. In addition, the reconstructed orbit 
indicates that Terzan~5 always remained well confined within the bulge \citep{massari16}. 
All these facts strongly disfavor the possibility that Terzan~5 formed in a satellite galaxy later accreted by the Milky Way, 
and rather suggest that it is a genuine Galactic system, which formed and evolved in and with the bulge. 
More recently, \citet{ferraro21} discovered the presence of at least two distinct sub-populations with remarkably 
different ages also in Liller~1. One population has an age of 12 Gyr, and the other one possibly is as young as 1 Gyr. 
The oldest populations in Liller~1 and Terzan~5 are impressively similar, in agreement with the typical age of bulge 
GCs and most of bulge stars. This indicates strikingly that they both formed at the same cosmic epoch, 
thus likely from gas clouds with compatible chemistry. Indeed, the metallicity of the few giant stars investigated 
so far in Liller 1 is perfectly compatible with that of the old population in Terzan 5: [Fe/H]$\sim-0.3$ and 
[$\alpha$/Fe]$\sim+0.3$ \citep{origlia02}. 
An in-situ origin, instead of an accretion from outside, therefore seems to be well plausible also for Liller 1.

\citet{ferraro09,ferraro21} suggested that these stellar systems could be interpreted as the surviving relics 
of much more massive primordial structures that generated or contributed to the Galactic bulge formation, 
similarly to the giant clumps observed in the star-forming regions 
of high-redshift galaxies \citep{immeli04,carollo07,elmegreen08,genzel11,tacchella15, 
behrendt16}.
In fact, observations of high-z spiral galaxies often show bright UV clumps within them \citep{cowie95,vdb96, 
giavalisco96,elmegreen04,elmegreen05,shibuya16}, indicative of 
massive star-forming complexes. Although  many of these clumps are expected to migrate to the centre of the host galaxy, 
due to dynamical friction, 
and deposit their stars there, effectively building the bulge of these proto-galaxies \citep{immeli04,dekel09},
it is possible that a few of them survived the total disruption, producing stellar systems grossly appearing 
like massive stellar clusters \citep{bournaud16}. 
These fossil relics could have been extremely massive in the past thus being able 
to retain the iron-enriched ejecta of SNe explosions, 
possibly producing the sub-populations now observed in Terzan~5 and Liller~1.

Terzan~5 and Liller~1 would therefore belong to a new class of stellar systems named \citep{ferraro21} 
as {\it Bulge Fossil Fragments}: 
these are stellar aggregates with the appearance of massive GCs orbiting the Galactic bulge, 
formed at the epoch of the Galaxy assembling, and harboring in addition to the old population, 
a younger component. 

An alternative formation scenario for Terzan~5 and Liller~1 has been recently proposed by \citet{bastian21}.
The authors argue that their peculiar properties might be the result of an interaction
between an ancient massive stellar cluster and a giant molecular cloud (GMC) that took place a few Gyrs ago, thus forming
the younger population observed in both systems. Based on the hydro-dynamical simulations by \citet{mckenzie18}, 
they suggest that, as a result of such interaction, the massive cluster may be able to 
accrete gas from the GMC and eventually experience a new event of star formation.
In this scenario, the probability of encounters between clusters and GMCs
increases for massive systems on disky orbits, as in the case of Liller~1 and Terzan~5, but in general 
it remains pretty low. 
As these are rare events, only an handful of systems is expected to undergo this formation process within 
a Hubble time.
In addition, because of their rarity, stellar systems forming this way are expected to produce two discrete star-formation episodes. 
In this scenario, the metal poor ([Fe/H]$=-0.7$) sub-population of Terzan~5 is composed by captured field stars.   

In this ongoing and exciting discussion about the physical processes behind the formation of Terzan~5 and Liller~1, 
constraining the detailed star formation histories (SFHs) of their stellar populations 
represents an important piece of information.
Driven by this motivation, in this paper we perform the very first attempt to reconstruct 
the SFH of these complex stellar systems by using Liller~1 as a test-bench. 
In fact, the available photometric data-set for Liller~1 has a larger photometric completeness and 
the main sequence turn-off of the younger population is brighter than in Terzan~5.

The paper is structured as follows. Section~2 presents the available data-sets adopted for this study and it describes 
the main steps of the data-analysis. In Section~3, we described the relative proper motion analysis to separate likely 
cluster-members from field interlopers and in Section~4 we briefly describe the artificial star experiment construction.
Section~5 briefly describes the approach used to reconstruct the SFH of Liller~1 and reports on the main results.
In Section~6 we summarize our findings and we discuss them in the context of the formation scenarios for this system. 

\begin{figure}[t]
\centering \includegraphics[width=12.5cm]{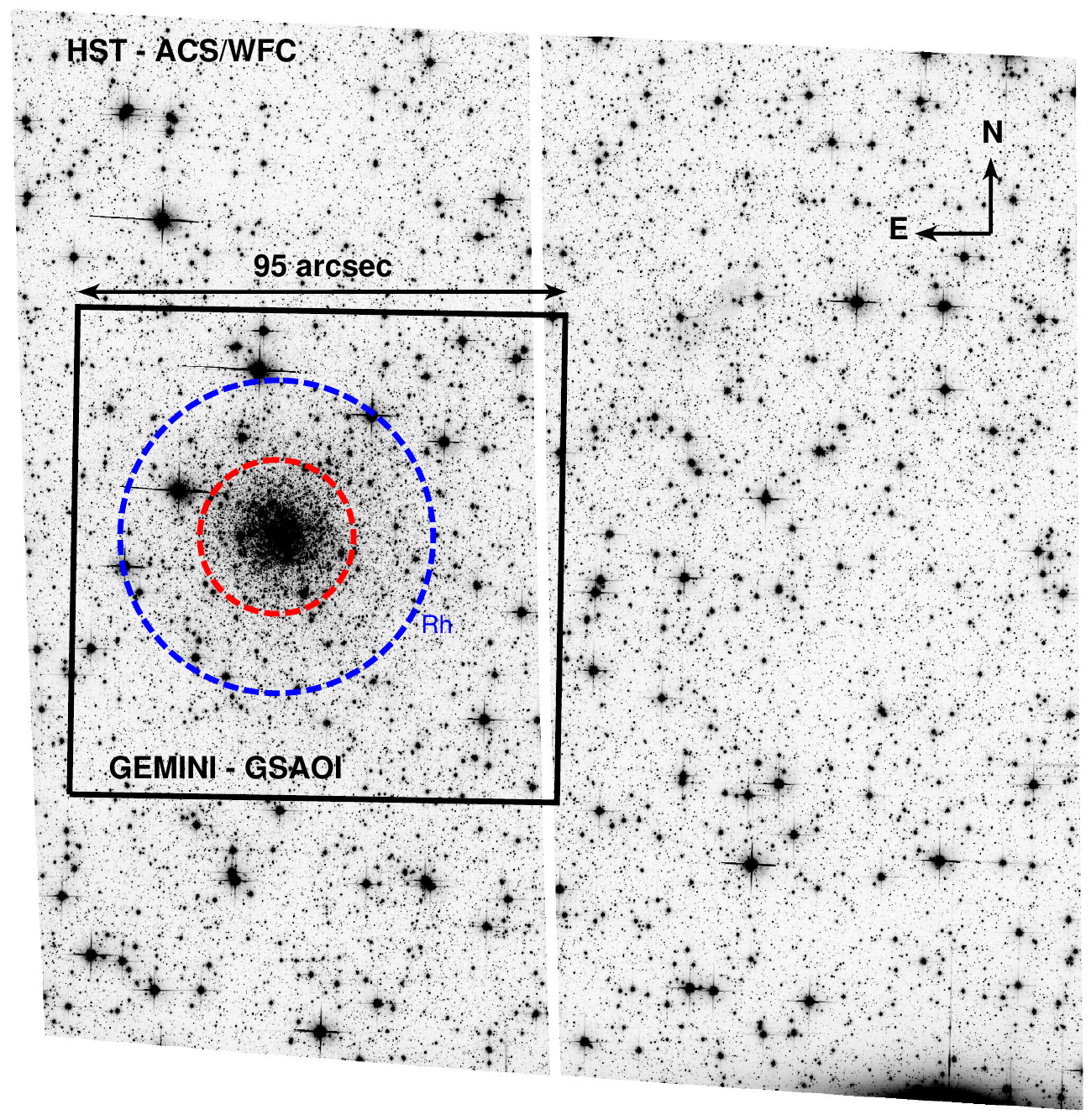}
\caption{HST ACS/WFC image of Liller~1 in the F814W filter. The field of view is 
$204\arcsec \times 204\arcsec$. The black box corresponds schematically to the GeMS-GSAOI
pointing ($95\arcsec \times 95\arcsec$). The region used for the SFH reconstruction is that beyond the red circle (having a radius of $15\arcsec$). 
The blue circle marks the half-mass radius of Liller 1 ($r_h=30.5\arcsec$, from \citealt{saracino15}). North is up, EAST is to the left.}
\label{map} 
\end{figure}

\section{Observations and data analysis}
The photometric dataset used in this study combines optical and IR high-resolution images. 
The optical database includes images acquired with the HST Wide Field Channel/Advanced Camera for Survey (ACS/WFC) through 
proposal GO 15231\dataset[10.17909/7zah-te68]{http://dx.doi.org/DOI: 10.17909/7zah-te68} (PI: Ferraro). It consists of 12 deep images obtained with the filters F606W 
(6 images with exposure time $t_{exp}=1300$sec) and F814W (6 images with $t_{exp}=600$). 
The ACS/WFC is composed of two twin chips, each of $4096\times2051$ pixels, separated by 
a gap of approximately 30 pixels. The pixel scale is $0.05\arcsec$ pixel$^{-1}$, therefore the resulting field of view 
(FOV) is $204\arcsec \times 204\arcsec$. As shown in Figure~1, the cluster is centered in chip $1$.
All images are dithered by a few pixels to allow a better subtraction of CCD defects, 
artifacts and false detections, and eventually a better sampling of the stellar point spread function (PSF). 

The IR dataset was obtained with the camera Gemini South Adaptive Optics Imager (GSAOI) assisted by the Gemini 
Multi-Conjugate Adaptive Optics System (GeMS) mounted at the 8 m Gemini South Telescope (Chile; Program ID: GS-2013-Q-23; PI: D. Geisler). GSAOI is equipped with a $2\times2$ mosaic of 
Rockwell HAWAII-2RG $2048\times2048$ pixels arrays with a resolution of $0.02$ pixel$^{-1}$ 
\citep{neichel14}. The central region of Liller 1 was sampled with a mosaic of multiple exposures acquired with a dithering pattern of a few arcseconds resulting in a global 
FOV of $95\arcsec \times 95\arcsec$ on the sky (Figure~1). Specifically, a total of 7 and 10 exposures were acquired in three nights from April 20 to May 24 2013 
in the J and K$_s$ bands, respectively, 
with $t_{exp}$ = 30 s each. 
This dataset was analyzed and presented by \citet{saracino15} in a first place. 
Here we re-analyzed only the 6 best-quality images in terms of delivered FWHM, 
Encircled Energy and Strehl ratio (see \citealt{dalessandro16}) for both the available filters.

\begin{figure}[t]
\centering \includegraphics[width=19.cm]{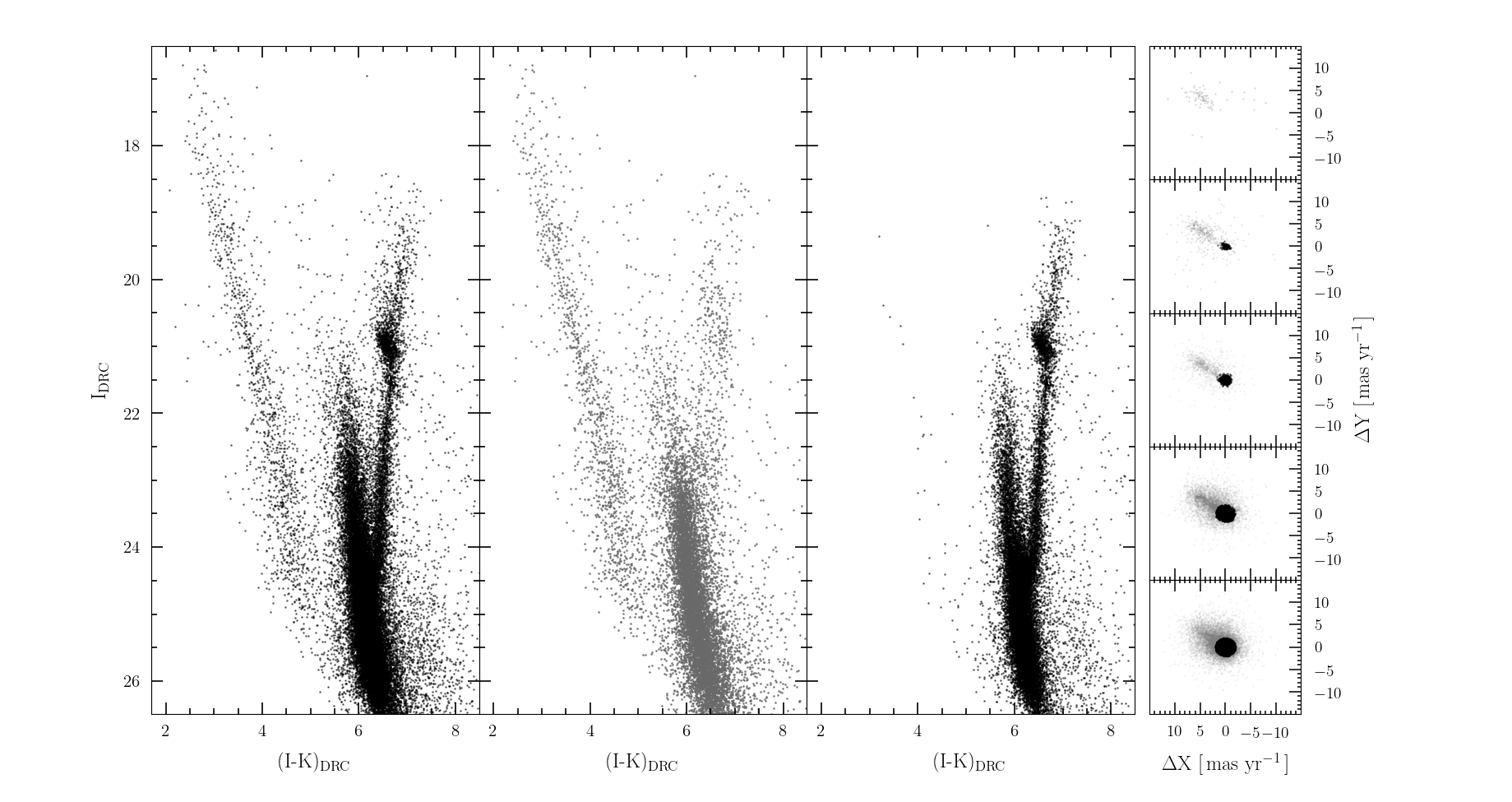}
\caption{Panel (a): differential reddening corrected (I,I-K) CMD of the stars in common between GEMINI and HST, and with measured PMs. 
Panel (b): CMD of the field population as selected from the VPDs as those with PMs not compatible with that of Liller~1. 
Panel (c): PM-cleaned CMD of Liller~1 obtained by using only the likely cluster members selected from the VPDs shown in the right-most column. 
Right-most column: VPDs of the measured stars divided in 5 bins of 2 mag each, starting at $I_{DRC}=16.5$. 
Black dots represent the stars selected as likely cluster members (see Section~3).
}
\label{cmd} 
\end{figure}

\label{subsec:fotometria}
The photometric reduction was carried out by performing PSF fitting 
in each chip of each image independently for both the IR and the optical data by using \texttt{DAOPHOTIV} \citep{stetson87}. 
The PSF has been modeled by selecting about 200 bright and isolated stars uniformly distributed in each chip, 
and by using the \texttt{DAOPHOTIV/PSF} routine. We allowed the PSF to vary within each chip following a cubic 
polynomial spatial variation. The PSF models thus obtained were then applied to the star-like sources detected at 
a $3\sigma$ level above the local background in all images by using \texttt{ALLSTAR}. 
We then created a master-list including stars detected in at least three HST F814W (that will be labeled hereafter as I) 
and GeMS K$_s$ (simply labeled as K) images. 
This choice was driven by the fact that both the F606W and J images are significantly shallower than the I and K ones because of 
the extremely high extinction  (E(B-V)$=4.52$; \citealt{ferraro21}) in the direction of Liller~1 and because of the reduced efficiency of the AO corrections 
in the J band. 
As done in previous works (e.g., \citealt{dalessandro18} and reference therein), the master list thus created
was used as input for \texttt{ALLFRAME} \citep{stetson94} and, at the corresponding positions of stars in the
master-list, a fit was forced in each frame of the two data-sets. 
For each star thus recovered, multiple magnitude estimates obtained in each chip were homogenized 
by using \texttt{DAOMATCH} and \texttt{DAOMASTER}, and their weighted mean and standard deviation were finally 
adopted as star magnitude and photometric error.
The obtained catalog includes 62,834 stars in total. Of them, 51,100 have both I and K magnitudes.

We reported the instrumental optical magnitudes onto the VEGAMAG photometric system by using the updated recipes 
and zero-points available in the HST web-sites\footnote{https://www.stsci.edu/hst/instrumentation/acs/data-analysis/zeropoints}. 
To calibrate the IR magnitudes, we used the stars in common between GeMS and the VISTA Variables 
in the V\'ia L\'actea (VVV) survey \citep{VVV}. The calibration zero-points were then set as the
difference between the J and K magnitudes in the two samples, after applying an iterative $3\sigma$-clipping algorithm.
The resulting (I-K, I) color-magnitude diagram (CMD), including the differential reddening corrections  (DRCs)
derived by \citet{pallanca21}, is shown in Figura~2 (left panel).

Instrumental coordinates (x,y) were first reported to the HST ACS/WFC reference frame. Then 
they were corrected for geometric distortions by using prescriptions by \citet{anderson10} and \citet{ubeda12} 
and were then transformed 
to the absolute coordinate $(\alpha,\delta)$ system by using stars in common with the publicly 
available early Gaia Data Release 3 (Gaia eDR3) catalog \citep{brown21}.
About 200 stars have been matched by using the cross-correlation tool \texttt{CataXcorr}, 
thus allowing a very precise determination of the stellar absolute positions. 
We note in passing that almost all the stars in common with the Gaia catalog are disc field stars. 
The resulting $1\sigma$ astrometric accuracy is $\sim 0.1\arcsec$.

\section{Proper motion analysis}
Figure~2 shows that, as expected given its position in the Galaxy, the evolutionary sequences of Liller~1 
are quite strongly contaminated by a significant fraction of field stars.
Very prominent is the blue sequence visible at I$_{DRC}<25$ and 
(I-K)$_{DRC}<5.5$ and likely populated by young disk stars. 

Unfortunately, Gaia eDR3 proper motion (PM) measures exist for only $\sim200$ stars in the system.
They are mostly distributed along the brightest portion of blue disk population, and,
because of their intrinsic faint magnitudes, 
the available astrometric measurements suffer from significant uncertainties (up to $\sim0.5$ mas yr$^{-1}$). 
Therefore, to clean the observed CMD from field star interlopers, we performed a relative PM analysis. 
In \citet{saracino20} we have shown that it is possible to efficiently use GeMS images in combination with the HST data to 
derive reliable relative PMs also in dense GCs (see also \citealt{fritz2017}).
PMs obtained with these data-sets
have been already used in \citet{ferraro21} and \citet{pallanca21}. Here we detail the approach adopted to derive them.

We used the ACS/WFC and the GeMS+GSAOI  data sets, which are separated by a temporal baseline of $\sim6.3$ yr, 
as first and second epochs,  respectively. 
PMs were derived by following the approach adopted by \citet{saracino20} (see also \citealt{dalessandro13}). 
The procedure consists in determining the displacement of the centroids of the stars measured in the two epochs, 
once a common coordinate reference frame is defined. The first step is to adopt a distortion-free reference frame, 
which we will call master frame hereafter. 
The master frame catalog contains stars measured in all 
the ACS/WFC I band single-exposures. Their coordinates were corrected for geometric distortions as described in 
Section 2.
To derive accurate transformations between the second epoch catalogs and the master catalog, 
we selected a sample of $\sim3300$ bona-fide stars having magnitude $20<I_{DRC}<24.5$ 
(corresponding approximately to magnitudes $13.0<K_{DRC}<18.5$), which we judged to be likely cluster members on the basis of 
their position in the CMD (i.e., stars distributed along the lower RGB, SGB and upper MS). 
We then applied a six-parameter linear transformation to report stars in the second epoch catalogs 
to the master coordinate reference frame, 
by using the stars in common.  
We treated each chip separately, in order to maximize the accuracy. 
Moreover, we carefully corrected the coordinates of stars in the second epoch catalogs for the important geometric distortions affecting 
the GSAOI camera by using the geometric distortions solutions published in \citet{dalessandro16}.

The mean position of a single star in each epoch (x$_m$, y$_m$) has been measured as the $3\sigma$ clipped mean 
position calculated from all the N individual single-frame measurements. The relative rms of 
the position residuals around the mean value divided by  $\sqrt N$ has been used as associated error ($\sigma$). 
Finally, the displacements are obtained as the difference of the positions (x$_m$, y$_m$) between the two epochs for 
all the stars in common. The error associated with the displacement is the combination of the errors on the positions 
of the two epochs. 
We should stress here that the first epoch is by construction in the same coordinate reference frame as the master catalog.
The relative PMs ($\mu_x, \mu_y$) are finally determined by measuring the difference of the mean x and y positions of the 
same stars in the two epochs, divided by their temporal baseline $\Delta T = 6.3$ yr. 
Such displacements are in units of pixels yr$^{-1}$. 
We then iterated this procedure a few times by removing likely non-member stars from the master reference frame based 
on the preliminary PMs obtained in the previous iterations. 
The convergence is assumed when the number of reference stars that undergoes this selection changes by less than $\sim 10\%$ 
between two subsequent steps. 
At the end we derived relative PMs for 35,761 stars in the area where the HST and GeMS datasets overlap.
The rightmost panels of Figure~2 show the derived vector point diagrams (VPDs) at different magnitude levels. 
As expected, the PM distributions get broader for increasing magnitudes because of the increasing 
uncertainties of the centroid positions of faint stars. 

To build a clean sample of stars with a high membership probability, we defined for each magnitude bin 
a different fiducial VPD region centered on (0,0). The fiducial regions have 
radii of $3\times\sigma$, where $\sigma$ is the average PM error in that magnitude bin.  
The sample of stars thus selected corresponds to the observational catalog we will be using for the following analysis. 
The resulting CMD is shown in Figure~2 panel c), while panel b) shows the CMD of likely field stars that have been excluded.

Because of the partial overlap between the motion of Liller~1 and that of Bulge and disk stellar populations in this region of the Galaxy, a possible residual contamination is expected also in the PM-cleaned CMDs. 
As a sanity check, we estimated the fraction of not accounted field population by using the `Besan\c{c}on model of stellar population synthesis of the Galaxy' \citep{robin03}.
We retrieved the output of the model in the direction of Liller~1 for a solid angle 
corresponding to the area sampled by the Gemini data ($95\arcsec \times 95\arcsec$; Section~2).
Following the PM selection described before, we selected stars moving with the same 
transverse velocity components as Liller~1 (($\mu_{\alpha} cos\delta,\mu_{\delta})=(-5.403,-7.431)$ mas/yr from \citealt{vasiliev21}) within a tolerance range corresponding to the 
$3\times \sigma$ selection adopted in Figure~2. 
In the Besan\c{c}on simulation we find 2635 residual contaminating stars out 
of 22861 Liller~1 observed member stars (i.e. selected as members based on our analysis).  
If we then limit this comparison to the turn-off 
region ($24<I_{DRC}<25$), we find $\sim490$ residual contaminating stars in the Besan\c{c}on 
model out of 7447 likely member stars, thus yielding a residual contamination of $\sim6\%$.
We therefore expect a negligible impact of such a small residual contamination on the following analysis.

\begin{figure}[t]
\centering 
\includegraphics[scale=0.37]{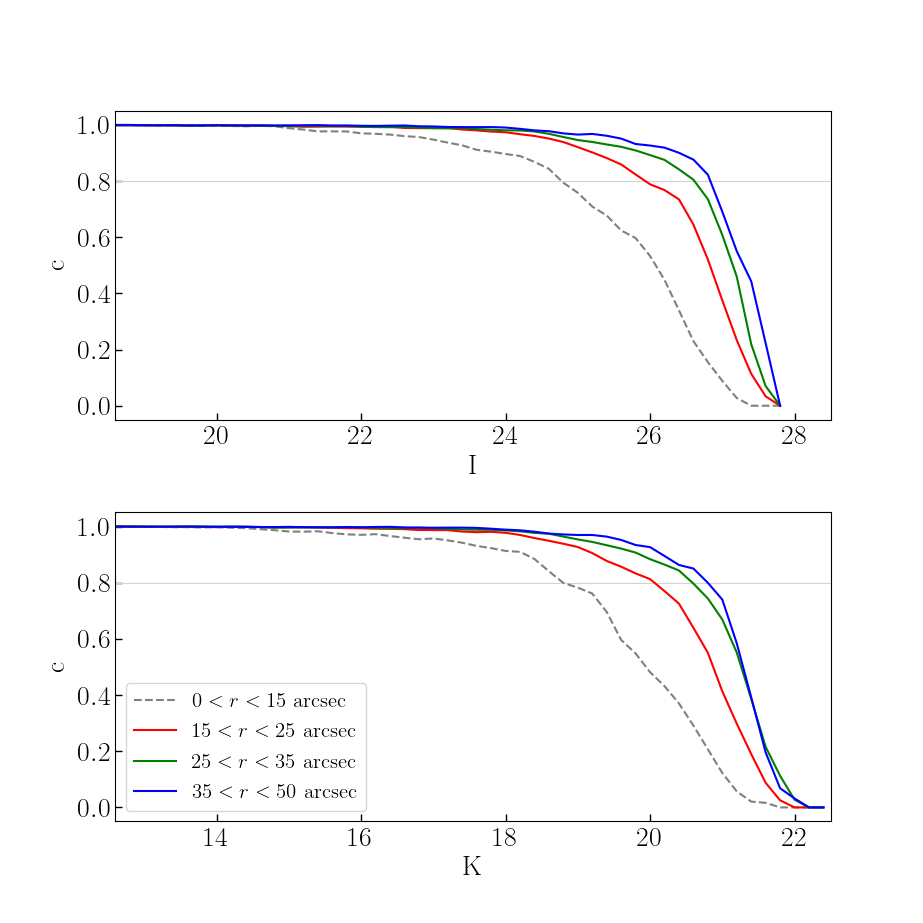}
\includegraphics[scale=0.35]{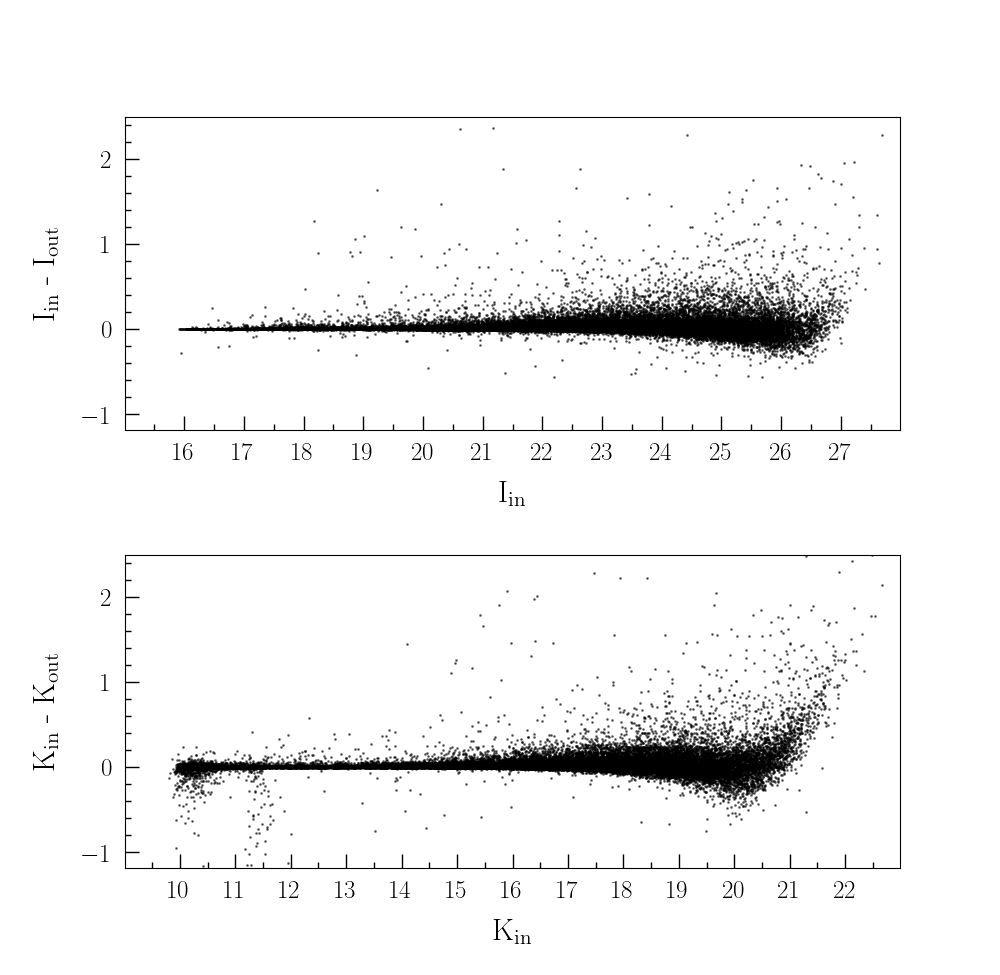}
\caption{Left panels: Completeness curves as a function of the I and K obtained in three concentric annuli at 
different cluster-centric distances. The horizontal grey lines marks the 80$\%$ completeness level for reference.
Right panels: differences between the input and output magnitudes in the I and K bands (upper and bottom panels respectively) for the artificial stars recovered 
by the photometric analysis.
}
\label{compl} 
\end{figure}

\begin{figure}[t]
\centering 
\includegraphics[scale=0.5]{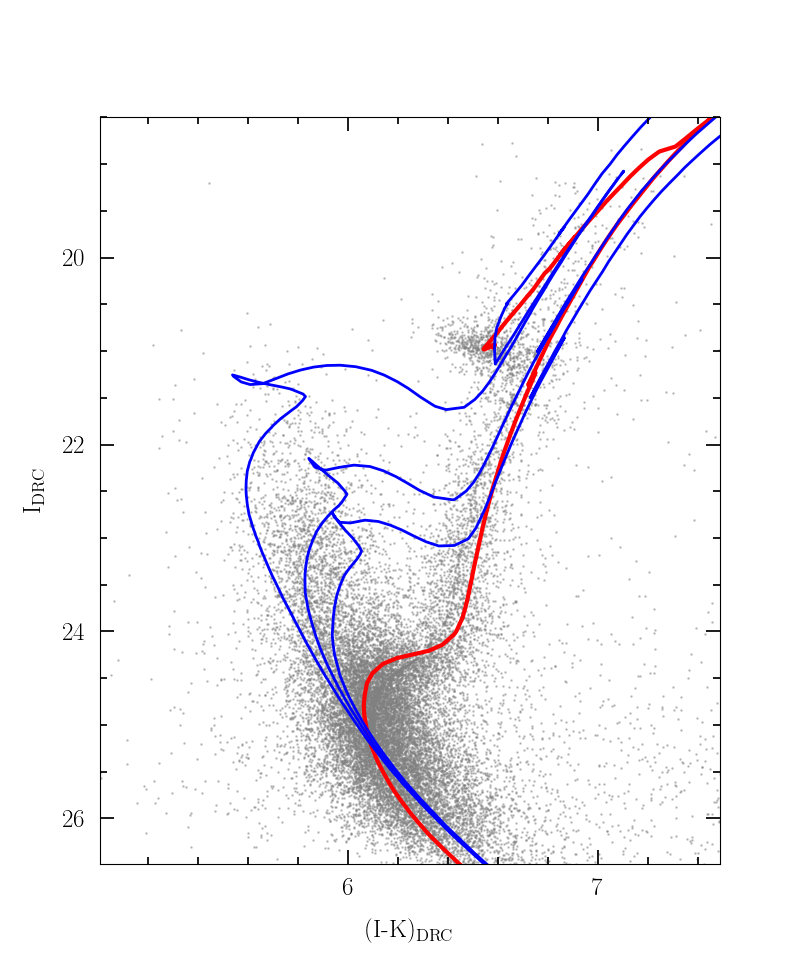}
\caption{Sample of isochrones over-plotted to the differential reddening corrected and PM selected CMD of 
Liller1 (grey dots): in red the 12 Gyr old isochrone with [M/H]=-0.3 that nicely reproduces 
the old stellar population, in blue three  young isochrones (of 1, 2 and 3 Gyr,  from top to bottom) 
at larger metallicity [M/H]=+0.3, which are needed to reproduce the locus occupied by the young population. }
\label{cmd_iso} 
\end{figure}

\section{Artificial Star Experiments}
For the reconstruction of the star formation history (SFH) of Liller~1, 
an accurate determination of photometric errors and incompleteness of the adopted data is needed. 
To this aim, we performed extensive artificial star experiments following the approach described in \citet{dalessandro15} 
with some ad-hoc modifications that are detailed below.
We generated a catalog of simulated stars with an I-band input magnitude ($I_{in}$) extracted from a flat luminosity 
function (LF) extrapolated beyond the observed limiting magnitude in the (I-K, I) observed CMD. 
Then, to each star extracted from the LF, 
we assigned an input K magnitude ($K_{in}$) from a randomly extracted value of color in the 
interval $2<$(I-K)$<9$ to homogeneously 
sample the range of magnitudes and colors occupied by stars in the observed CMD.
We note that this is a critical aspect for the present analysis and it differs from the approach adopted 
in previous papers of our group 
and from what was done in \citet{ferraro21} where artificial stars
follow the mean loci of the main evolutionary sequences defined by the observations. 
 
We then added artificial stars to real images adopting the same PSF models resulting from the photometric analysis
by means of the \texttt{DAOPHOTIV/ADDSTAR} package. 
Artificial stars were placed into the images in such a way they follow the stellar density profile 
obtained by \citet{saracino15} for Liller~1. 
To avoid artificial crowding,
stars were placed in a regular grid composed by 20 $\times$ 20 pixel cells (which correspond to $\sim$10 times 
the stellar FWHM in the HST images), where only one artificial star was allowed to lie. 
In this way, for each run we could add a maximum number of $\sim$ 5000 artificial stars. 
We repeated this analysis several times to reach a final simulated catalog composed by 
more than 350,000 artificial stars, which is about 6 times larger than the number of observed stars.  
For each iteration, we performed the same photometric analysis described in Section~2.
Those stars recovered after the photometric analysis have also values for $I_{out}$ and $K_{out}$.

The completeness is then defined as the ratio between the number of recovered artificial stars and that 
of the injected ones. 
Specifically, a star was not counted if it was not recovered during the photometric reduction, 
it was not included in the common ACS/WFC and GeMS field of views or it had a 0.75 brighter magnitude 
than its input to exclude cases 
where an artificial star is placed in the same spatial position as a real one with 
the same or brighter magnitude.
The left panels of Figure~3 show the completeness curves as a function of both the I and K bands and obtained for four concentric annuli at  
different cluster-centric distances.
The right panels show the distributions of the ($I_{in}$-$I_{out}$) and ($K_{in}$-$K_{out}$) as a function of the input I and K bands, 
respectively.
These distributions are used to quantify the photometric errors and the fraction of possible blends 
as a function of the magnitude and color.

\begin{figure}[t]
\centering \includegraphics[width=12.5cm]{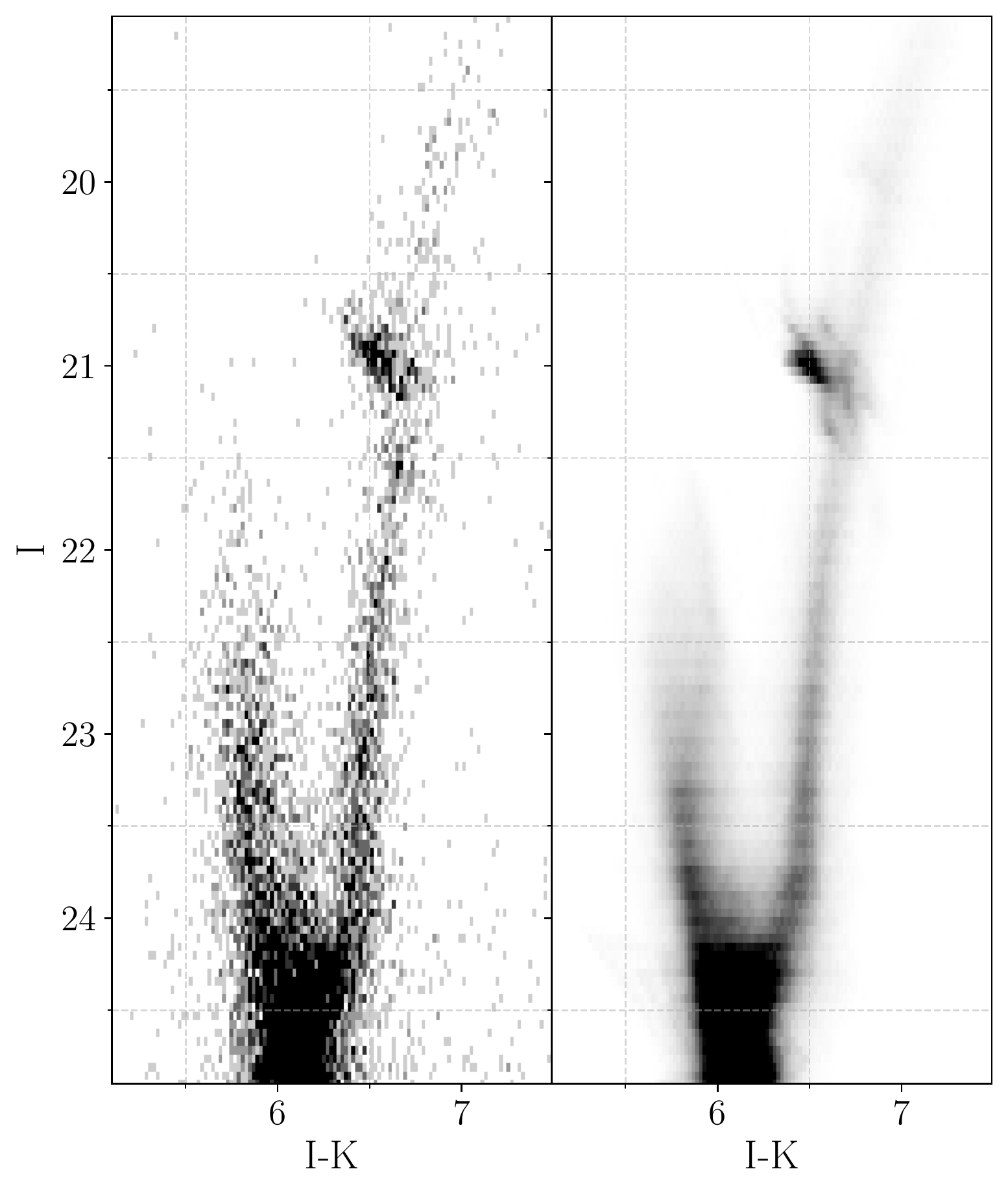}
\caption{Hess diagrams for the observed CMD (left) and the best-fit synthetic CMD (right).
}
\label{2cmds} 
\end{figure}

\begin{figure}[t]
\centering \includegraphics[width=12.5cm]{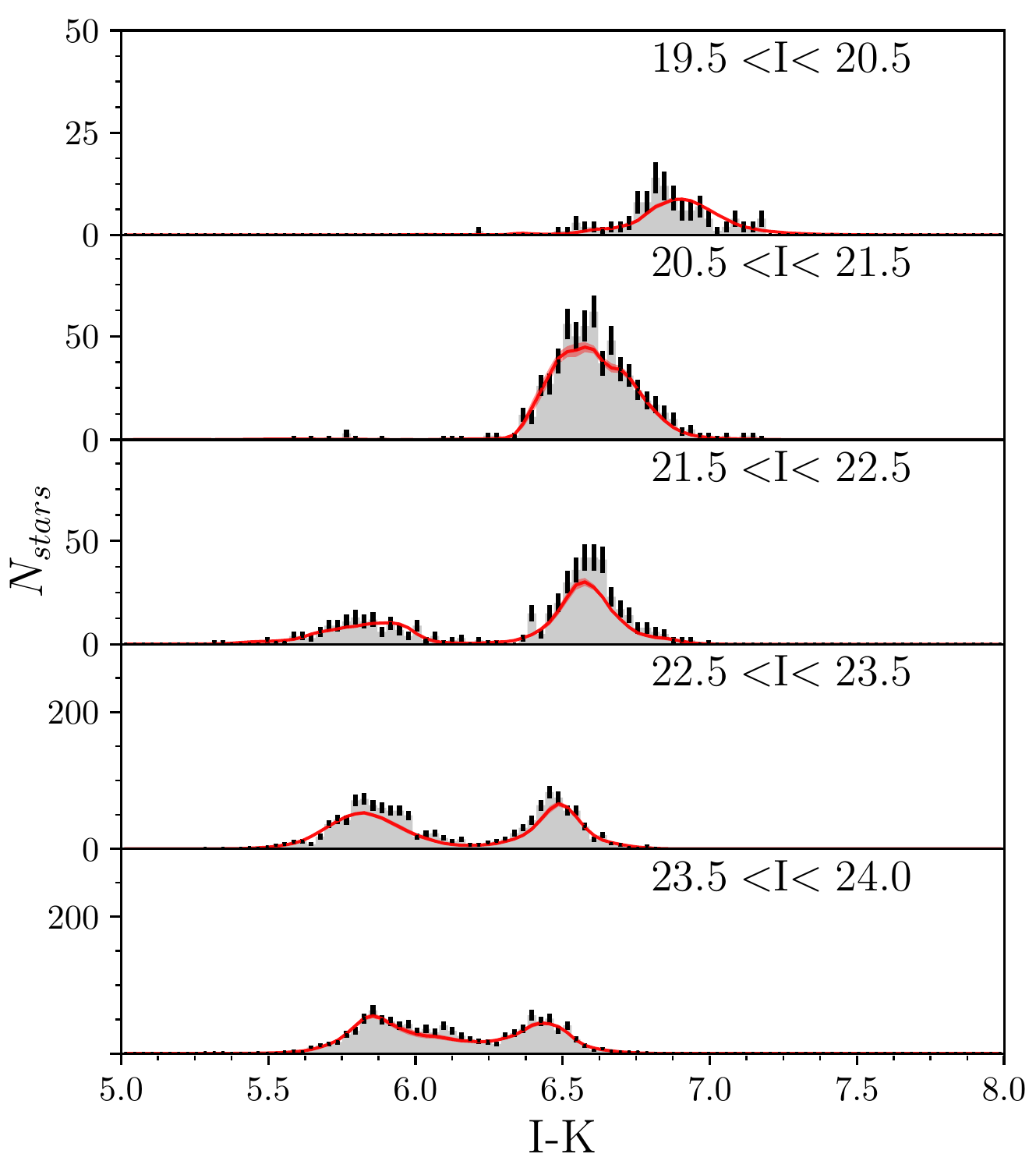}
\caption{Observed color distributions of Liller 1 in four magnitude bins (grey shaded areas), compared 
to those obtained from our best-fit SFH solution (red lines). 
The black vertical segments correspond to the Poissonian error in each color bin, while the uncertainty 
in the synthetic distribution is marked by the red-shaded areas. }
\label{CF} 
\end{figure}

\section{The Star Formation History analysis}
\subsection{The method}
Figure~\ref{cmd_iso} shows a zoomed view of the DRC and PM-selected CMD of Liller~1.
To guide the eye we super-imposed a set of isochrones of different ages 
retrieved from the PARSEC database \citep{bressan12}.  
We adopted a distance modulus (m-M)$_0=14.65$ and extinction E(B-V)$=4.52$ \citep{ferraro21} 
to match the models to the observations.  For the extinction law we adopt
\cite{cardelli89} with a total-to-selective extinction value
R$_{V}=2.5$, as suggested by \citet{pallanca21}.
As already found by \citet{ferraro21}, such a simple comparison shows that the bulk population of Liller~1 
is grossly reproduced by a model with
sub-solar metallicity ([Fe/H]$\sim-0.3$) with an age $t=12.0 \pm 1.5$ Gyr and by a more metal-rich 
([Fe/H]$\sim+0.2$) and significantly younger 
population with age ranging from $\sim3$ to $\sim1$ Gyr.
These results broadly indicate that Liller~1 might have experienced at
least two episodes of star formation, with the second
one possibly having formed stars from a more metal-rich gas.

Here we attempt to move a step forward 
and to characterize in detail the stellar populations observed in Liller~1
and their SFH.

The SFH was determined using the population synthesis routine Star Formation Evolution Recovery Algorithm 
(SFERA; \citealt{cignoni15}), 
applied to the differential reddening and PM-cleaned CMD shown in Figure~\ref{cmd_iso}.
SFERA employs a synthetic CMD method, along the lines pioneered by \citet{tosi91}. 
We provide here only a short description of the SFERA's approach, 
and we refer the reader to  \citet{cignoni15,cignoni16,cignoni18} for further details. 
Briefly, SFERA builds a library of synthetic CMDs starting from a set of
theoretical models spanning a wide range of parameters including age, metallicity, stellar 
initial mass function (IMF), binary
fraction and a range of distance and extinction values. The
model CMDs are then convolved with observational errors and biases as
measured from artificial star tests. Finally, SFERA employs a maximum
likelihood statistics to compare synthetic and observed CMDs and
recover the most likely SFH.

\begin{figure}[t]
\centering \includegraphics[width=13.5cm]{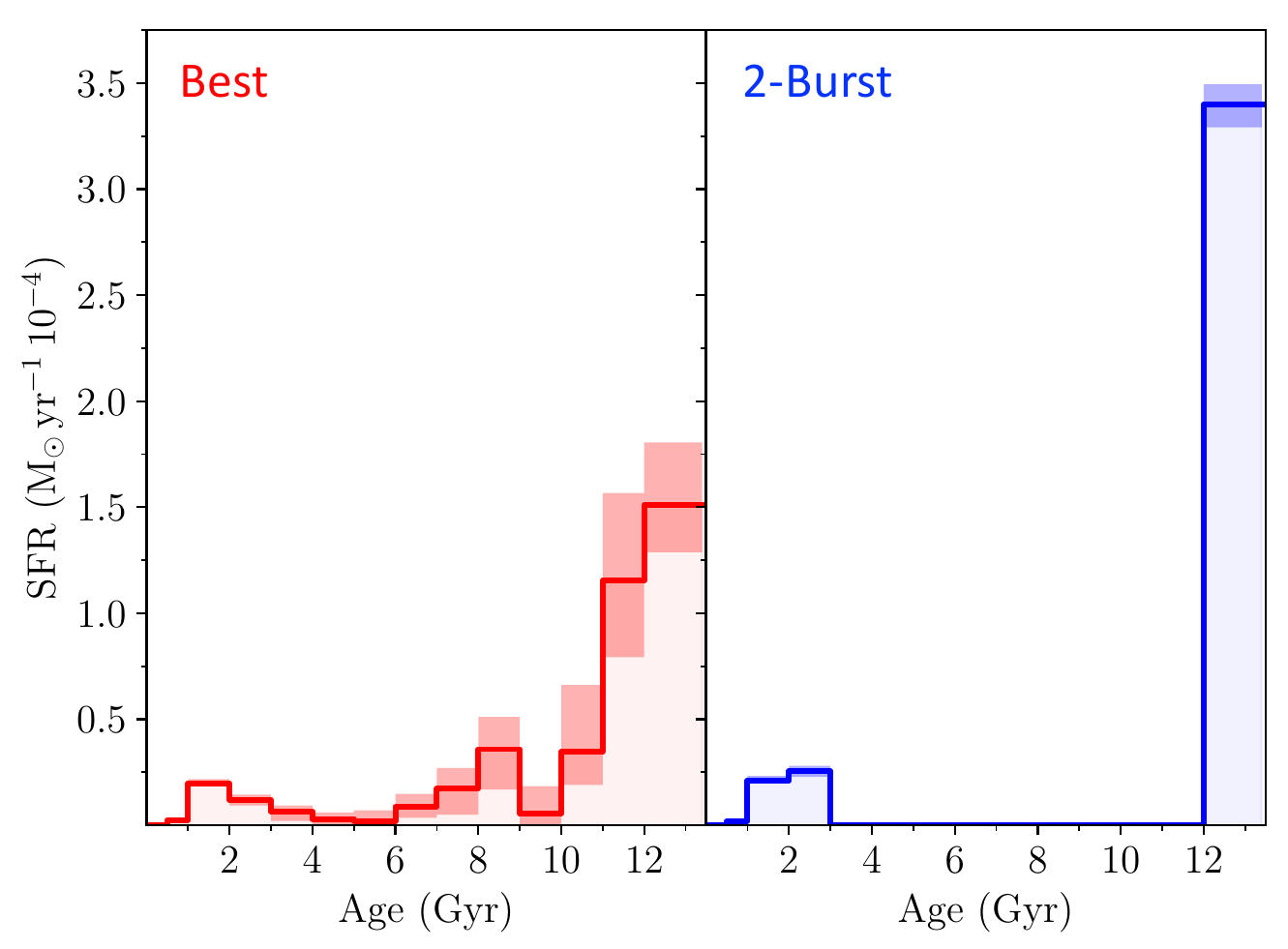}
\caption{Best (red) and `2-burst' (blue) SFHs of Liller~1. The shaded areas define the regions between the 5th and
  95th percentile values.}
\label{sfh} 
\end{figure}

Synthetic CMDs are generated by using isochrones from the PARSEC-COLIBRI 
\citep{bressan12,marigo2017} stellar evolution models.
In particular, all synthetic CMDs are Monte Carlo realizations of all
possible combinations of 14 equally spaced age bins, between the youngest ($2$ Myr) and the oldest
available isochrone ($13.4$ Gyr). Within each age bin, a range of
metallicities is allowed, from the lowest available metallicity,
$[M/H] = -2.0$, to the highest, $[M/H] =+0.3$\footnote{We adopt [M/H]$=\log(Z/Z_{\odot})$, with
$Z_{\odot}=0.0152$. We report also that [M/H]=[Fe/H] + $\lg(0.638*10^{[\alpha/Fe]} +0.362)$ form \citet{salaris93}.}. This metallicity range has been chosen to include 
 the available metallicity derivations of Liller~1 
and, at the same time, to secure a quite large tolerance for the iron abundance to vary. 
IR high-resolution spectroscopy has been obtained with NIRSPEC@Keck \citep{origlia02} 
for only an handful of stars. This analysis suggests 
an average value of about half solar metallicity, and some $\alpha$-enhancement ([$\alpha$/Fe]$\sim+0.3$). 
A \citet{kroupa01} initial mass function (IMF) between 0.1 and 300 $M_{\odot}$ is then used 
to fully populate the synthetic CMDs. The adoption of other commonly used IMFs, such as \citet{salpeter55} or \citet{chabrier03} IMFs, is expected not to impact significantly the main 
results of the present analysis. In fact, the quoted IMFs do not differ significantly for stellar masses larger than $\sim~1 M_{\odot}$, which is approximately the mass regime we focus on.
Unresolved binaries are also considered and $30\%$ of synthetic stars are coupled with a stellar
companion sampled from the same IMF. 

Distance and foreground extinction are also free parameters, whereas
differential extinction is not considered as it has been corrected star by star as discussed
above in the observations.

To properly compare synthetic CMDs with the observed ones, 
we need to convolve theoretical models with all the observed sources of uncertainties, such as 
photometric errors, blends and photometric incompleteness.
We account for observational errors and incompleteness by
smearing the synthetic CMDs with the color and magnitude distribution 
of errors and completeness derived from our artificial star tests (Section~4).  

To derive the SFH and identify the best-fit model, SFERA first
constructs a Hess diagram for the observed data, and then attempts to
match this by linearly combining Hess diagrams from our library of
synthetic CMDs. 
The best combination of model Hess diagrams is then obtained minimizing a
Poissonian likelihood (function of the data-model residuals) by means
of a hybrid-genetic algorithm.

The severe crowding conditions of the innermost regions of Liller 1 
hindered the possibility to effectively use the turn-off region for the entire available FOV. 
For this reason we decided to
limit the SFH analysis to stars at cluster-centric distances larger than $15\arcsec$, 
where the photometric completeness is generally larger than $80\%$ for $I\sim 25$ (see Figure~\ref{compl}). We estimate that with such a selection we sample $\sim40\%$ 
of the total mass of Liller~1.

\subsection{The best-fit SFH}
SFERA estimated a best fit extinction corrected distance modulus of 14.65 and foreground
reddening of $E(B-V) = 4.52$.  These values nicely compare with
those previously derived by \citet{ferraro21} and 
\citet{pallanca21}.

\begin{figure}[t]
\centering \includegraphics[width=11.5cm]{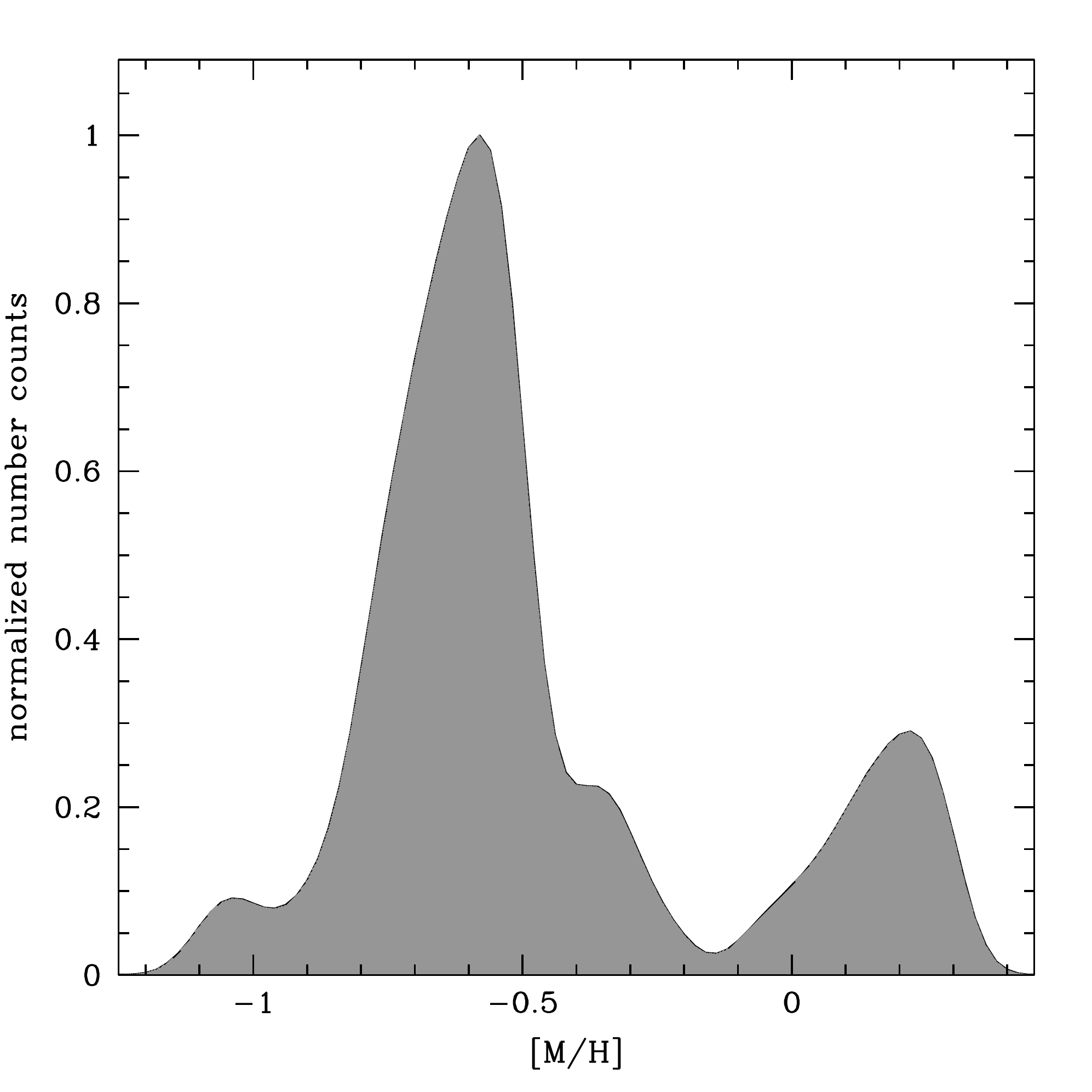}
\caption{Generalized histogram of the metallicity distribution of Liller 1 as constrained from the best-fit SFH.}
\label{met} 
\end{figure}

In Figure~\ref{2cmds} we compare the observed and the best-fit synthetic CMDs (left and right panels respectively) 
by showing their Hess diagrams for a first visual check.  
At a qualitative analysis, the synthetic CMD indeed nicely reproduces the main 
features of the observed one.
In fact, it matches the observed turn-off, the color and magnitude extension of the blue plume, 
the RGB color width as well as the color extension 
and inclination of the red clump.

\begin{figure}[t]
\centering \includegraphics[width=13.5cm]{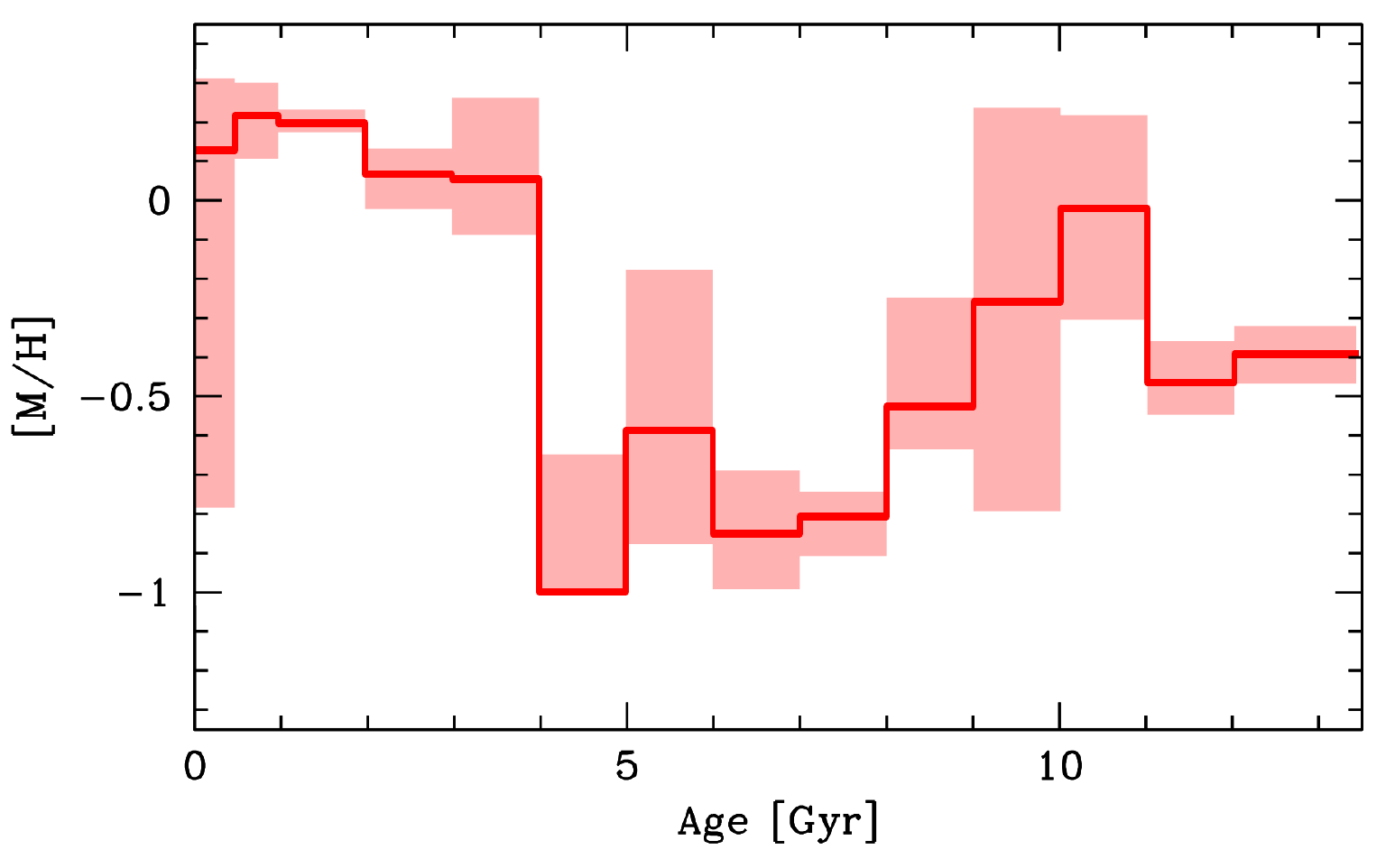}
\caption{Age-metallicity distribution as recovered from the best-fit SFH. The shaded areas corresponds to the 5th and 95th percentile values.}
\label{zt} 
\end{figure}

\begin{figure}[t]
\centering \includegraphics[width=15.5cm]{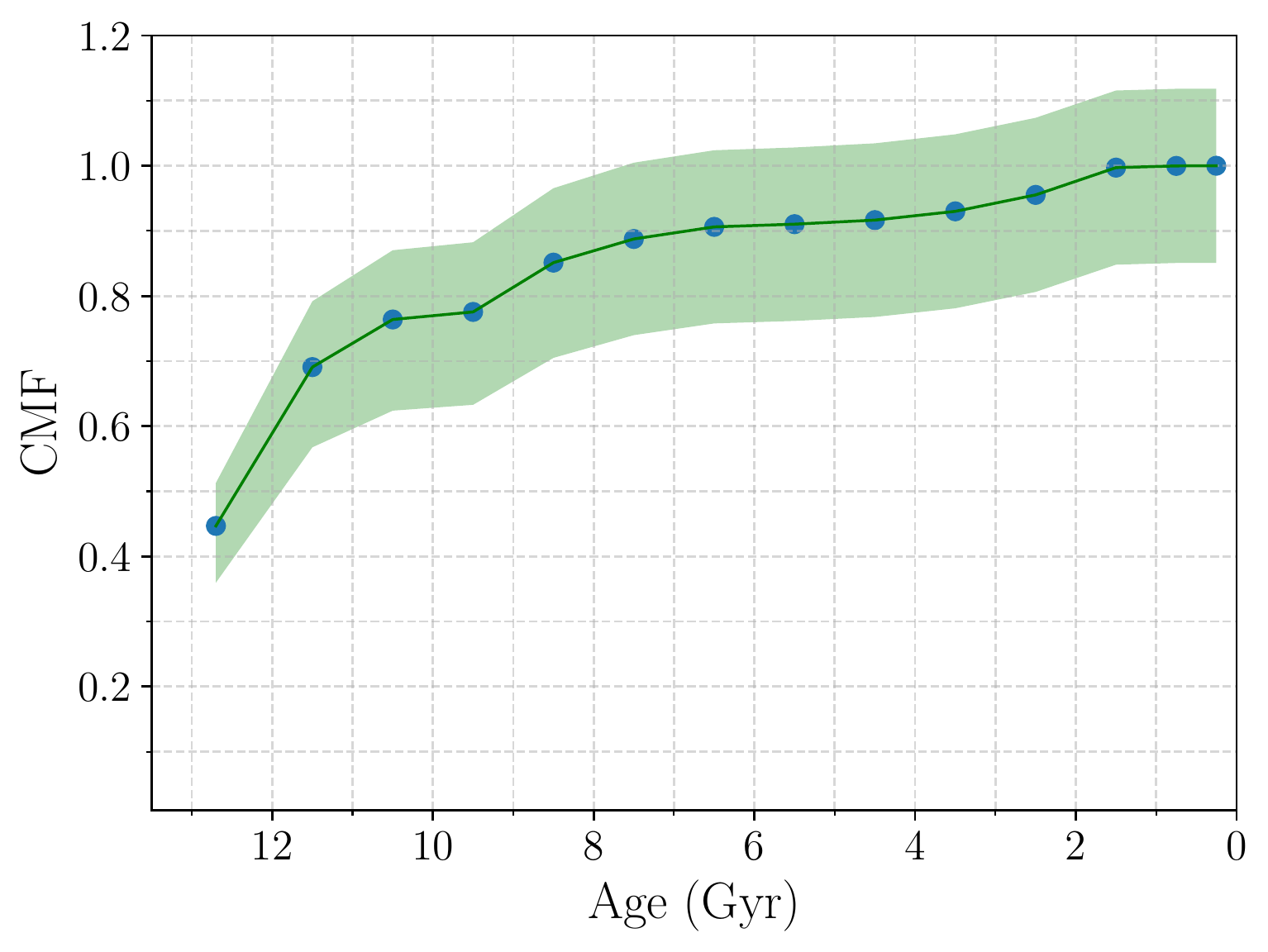}
\caption{Cumulative mass distribution for the recovered SFH.}
\label{cd} 
\end{figure}

For a more quantitative analysis, in Figure \ref{CF} we directly compare the 
color distributions obtained from the observed and the best-fit synthetic CMDs
in four magnitude bins. 
The best-fit synthetic CMD reproduces fairly well the mean color and the width of all the main observed 
evolutionary sequences, even if in the mid/bright portion of the RGB and along the young MS 
($21.5<I<23.5$) the best-fit synthetic CMD tends to slightly under-predict the number of observed stars.
This difference can be ascribed to residual contamination
from bulge + disk stars along the line of sight (see Section~3) 
and to a slight overestimate of the photometric completeness at the turn-off
magnitude level of the old population (I$\sim25$), which then produces smaller number counts at 
brighter magnitudes in the synthetic CMDs.

\begin{figure}[t]
\centering \includegraphics[width=13.5cm]{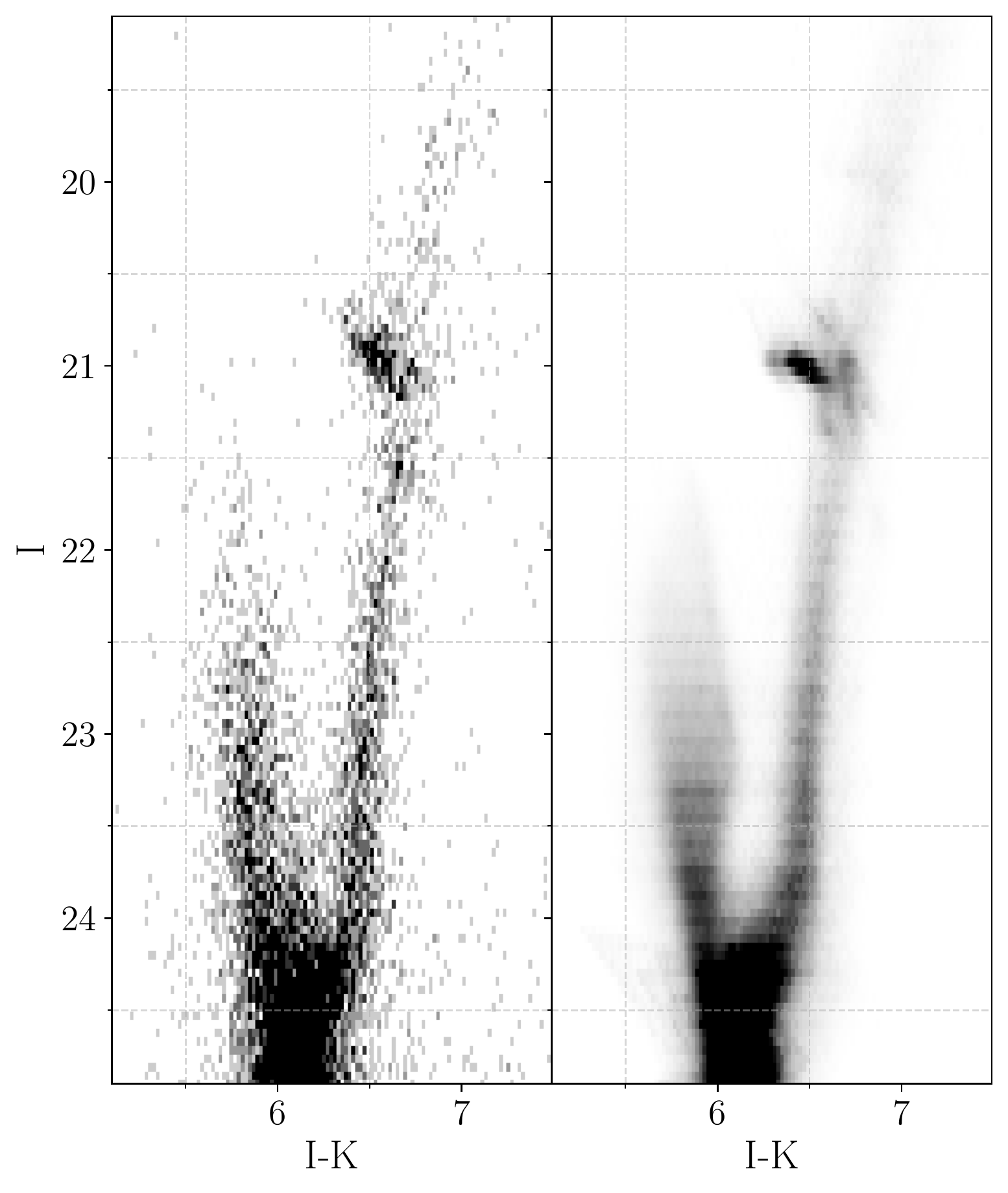}
\caption{As in Figure 5, but now for the `2-burst' SFH solution.}
\label{2b} 
\end{figure}

\begin{figure}[t]
\centering \includegraphics[width=12.5cm]{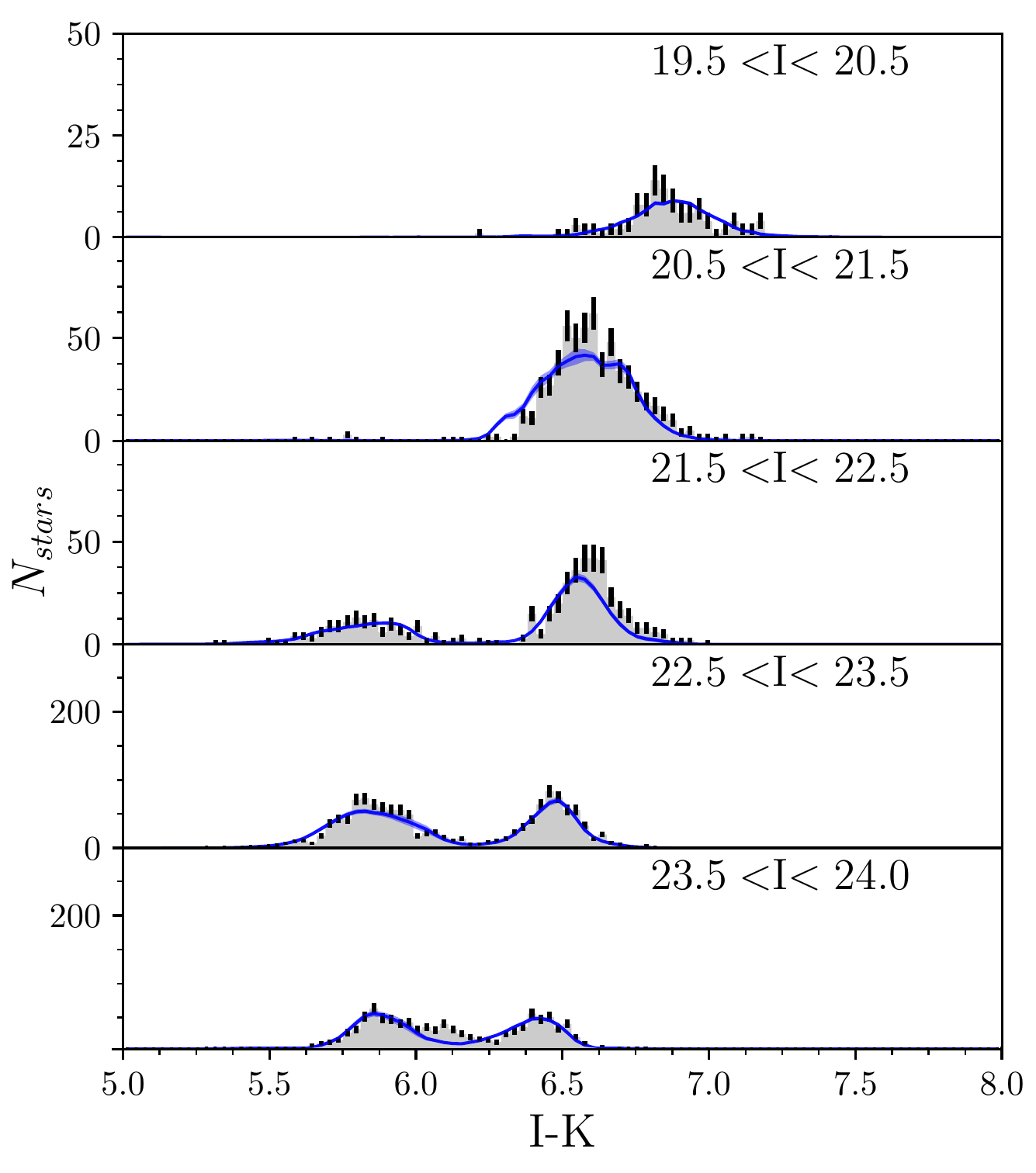}
\caption{As in Figure~6, but now for the `2-burst' SFH solution.}
\label{col_func_2b} 
\end{figure}

\begin{figure}[t]
\centering \includegraphics[width=15.cm]{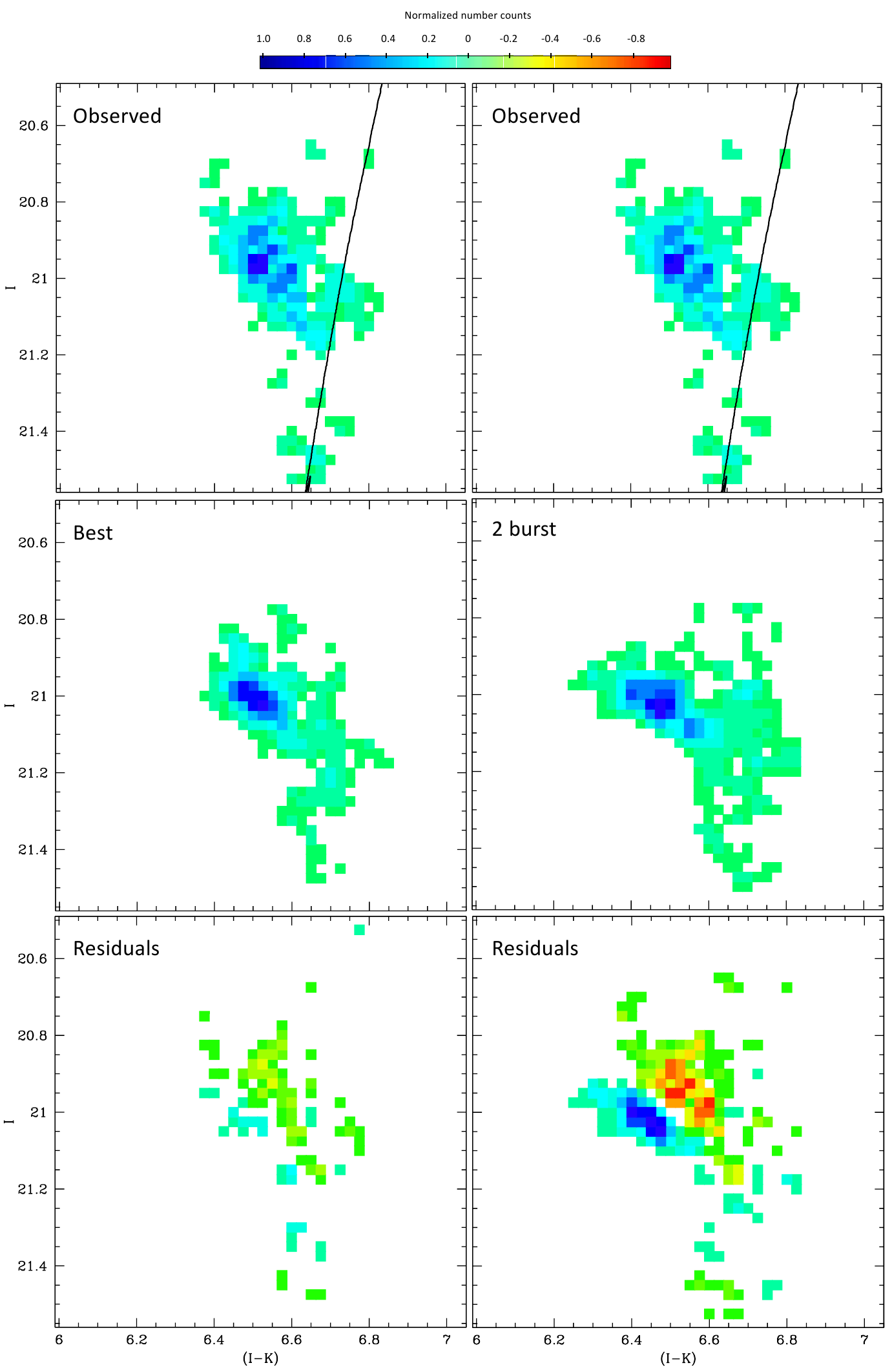}
\caption{
Upper panels: 2D distribution of the observed CMD density number counts in the red clump region.
Middle panels: distributions of the synthetic stars as obtained from the best-fit and 2 burst SFH 
solutions. Lower panels: CMD distribution of the difference between the number of observed and synthetic stars.}
\label{RCs} 
\end{figure}

The resulting best-fit SFH is shown in red in Figure~\ref{sfh}. 
The boundaries of the light-red shaded rectangles correspond to the 5th and 95th percentile values of the distribution of all the synthetic CMDs produced by SFERA and 
compared with the observations. 
The first key result emerging from Figure~\ref{sfh} is that Liller~1 
has been active in forming stars over the entire Hubble time. 
More specifically, three SF episodes are clearly detected: 1) a dominant one,
occurred 12-13 Gyr ago with a tail extending for up to $\sim3$ Gyr; 2) an intermediate burst, between 6 and 9 Gyr ago; 3)
a recent one, between 1 and 3 Gyr ago.  In addition, our analysis shows robustly 
that about 1 Gyr ago Liller~1 stopped forming stars. It is worth stressing here that we can exclude this 
result arises because of saturation issues. In fact, saturation appears to become important at I$_{DRC}<19$, which is more than 3 magnitudes brighter 
than the turn-off magnitude of the 1 Gyr population.

In the best-fit recovered SFH, we find a broadly bi-modal metallicity distribution (see Figure~\ref{met}). The stellar metallicity mostly fluctuates in the range $-0.8< [M/H]<-0.4$ dex at the older epochs, then  
it increases with time and the metallicity of the younger populations (t$<3$ Gyr) 
peaks at [M/H]$\sim+0.2$ (Figure~\ref{zt}).

The cumulative mass distribution related to such a SFH is presented in Figure \ref{cd}. 
It shows that more than $\sim70\%$ of the total mass was produced during the first event of SF ($t>10$Gyr), 
thus confirming that the bulk population of Liller~1 is old \citep{ferraro21}.
The remaining $\sim30\%$ is then almost equally split between the second and third SF events, 
with the latter producing no more than 
$\sim10\%$ of the total mass. We should recall at this point that the fraction of mass produced in the last 3-4 Gyr
might be underestimated because of the radial selection adopted to perform the analysis. 
In fact, the younger stellar population of Liller~1 is more centrally 
concentrated than the older ones and the number ratios between the young and old populations, 
is $0.98\pm0.04$ and $0.66\pm0.02$ for distances smaller and larger than $15\arcsec$, respectively (see \citealt{ferraro21}). 
We also stress here that, while Blue Straggler Stars overlap with the distribution of 
young stars in the CMD, they have only a limited impact on the recovered SFH of the young component ($<3$ Gyr) as they are expected to represent the $5-10\%$ of the total number of the blue plume population (see \citealt{ferraro21} for details about the number counts.)   

\begin{figure}[t]
\centering \includegraphics[width=16.5cm]{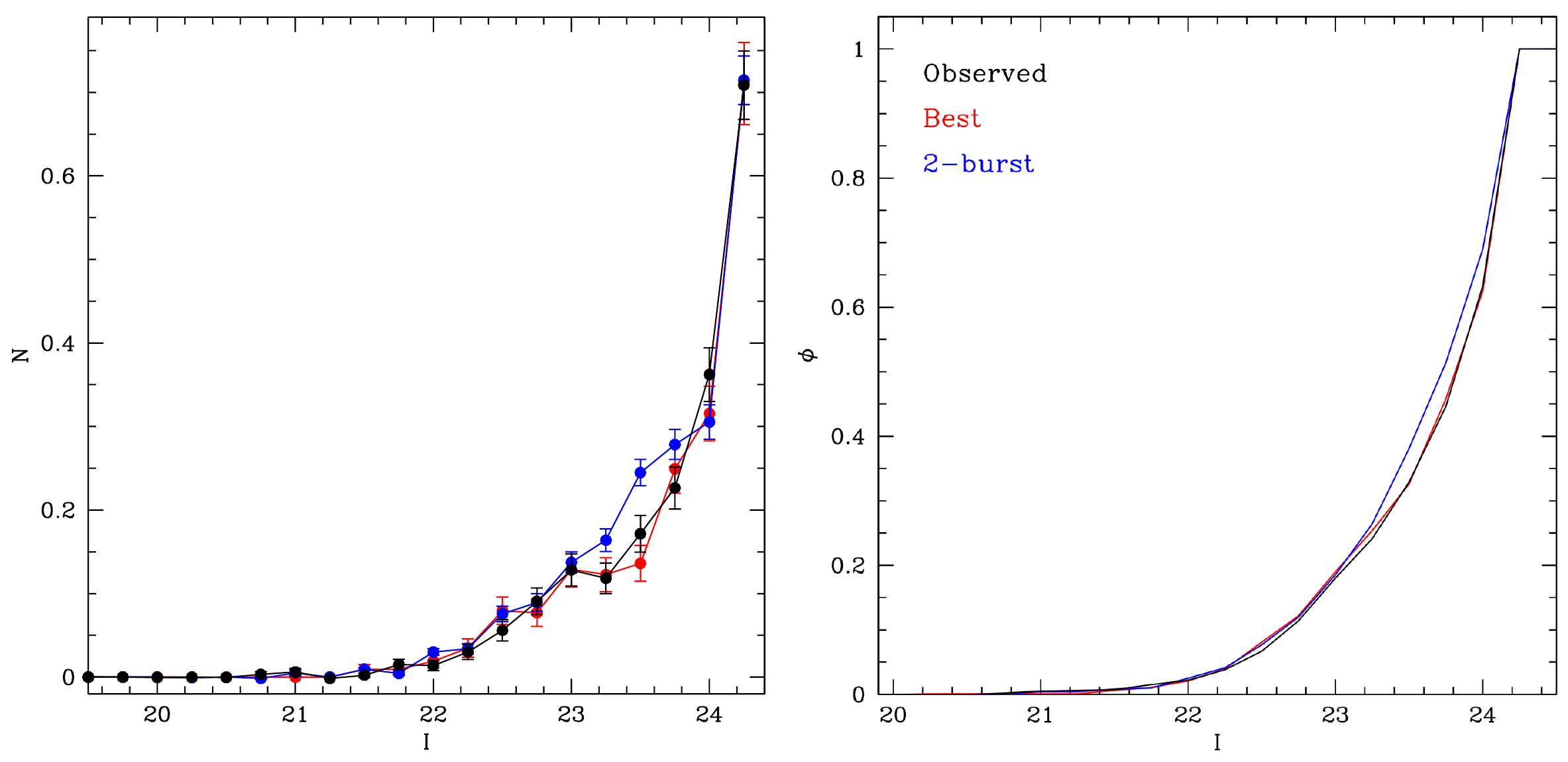}
\caption{Upper panel: Normalized luminosity functions of the young MS population (I$<24$, I-K$<6.3$).
In black is the observed distribution, while the red and blue curves refer to the best and 2 burst SFH solutions, respectively.
Lower panel: the difference between the best and 2 burst LFs with respect to the observed are shown. }
\label{lf} 
\end{figure}

\subsection{Alternative SFH solutions}
We tested the possible impact of poorly modeled observational uncertainties 
on the recovered SFH, and in particular on the results related to the older component where they can be more important because of the poorer ability to effectively distinguish populations 
with relatively small age differences.
More specifically, we checked whether the old and the
intermediate SF episodes can be safely considered as two separate events and, whether 
the resulting non negligible duration ($\sim3$ Gyr) of the old SF episode is real. 
To this aim, we re-performed the SFH analysis by imposing a `2-burst' solution in which the first SF episode 
is older than 12 Gyr and 
the more recent one is younger than 3 Gyr. 
The resulting SFH (blue slope) is compared with the previous one (red slope) in Figure~7.
The Hess diagram and the color numer count distribution corresponding to the newly derived SFH are shown in Figure~\ref{2b} and Figure~\ref{col_func_2b} respectively. 
Although a reasonable match between the observed and the synthetic CMDs seems to be achieved also in this case, 
a closer look shows that the two SFHs show non negligible differences in specific CMD regions.  

First, we note that when only two episodes of SF are assumed, the RC gets more 
horizontal than the observations and does not provide an adequate match to the observed color and magnitude distributions.
In Figure~\ref{RCs} we compare the CMD star density distributions in the RC region of the difference between the observations and the two SFH solutions.
Clearly, in the case of a SF with two discrete bursts the residuals gets larger and more structured.
The main responsible for such a mis-match is likely the lack of stars with ages in the range $7-12$ Gyr and likely more metal-rich than the old component.   

In addition, we note that the `2-burst' SFH solution provides a worse match also of the 
young population MS distribution (I$<24$, I-K$<6.3$).
In Figure~\ref{lf}, we compare the differential and cumulative luminosity functions (LFs) 
obtained from the observations (black line), with 
that resulting from the two SFH solutions (red and blue). The LF obtained with the best-fit SFH provides a better match of the observations. A KS test on the cumulative LF demonstrates that the `2-burst' LF is extracted from a different parent distribution than the observed one with a high degree of confidence (P$_{KS}\sim5\times10^{-3}$).  

Therefore, although we cannot completely exclude 
that the SFH of Liller~1 is bimodal and characterized by a very narrow old SF episode, 
we stress that a SFH history with three episodes, and 
characterized by a first event with possibly an extended tail, followed by an intermediate 
and a young SF episode, 
provides an overall better match to the observations.

\section{Summary and Conclusions}
Liller~1 and Terzan~5 are with no doubt among the most peculiar and interesting
stellar systems in the Milky Way, and their origin is still strongly debated.
Their present-day properties, such as their chemical abundance patterns and their similarity with those of 
the Galactic bulge \citep{ferraro09,origlia11}, their present-day orbits \citep{massari16}, as well as their 
age-metallicity-relations \citep{pfeffer21} would suggest they are genuine Galactic stellar systems and 
would exclude they are the remnants of an external massive stellar system accreted by the Milky Way, 
like a dwarf galaxy (e.g., \citealt{tolstoy09}) or its nuclear star cluster (e.g., \citealt{neumayer20}). 

Two main ideas have been put forward for their formation.
One suggests that Terzan~5 and Liller~1 are the remnants of primordial massive systems, likely produced by 
the fragmentation of an early disk,  that formed in situ and contributed to generate the bulge some 12 Gyr ago, 
the so-called Bulge Fossil Fragments \citep{ferraro09,ferraro21}. 
The detection of similar structures in the star forming regions 
of high-redshift galaxies confirms that such massive fragments likely existed at the epoch of the Milky Way assembly. 
The second scenario suggests that both Liller~1 and Terzan~5 are the result of a relatively recent encounter between
an old and massive GC formed in the Bulge and a GMC orbiting the Galactic disk \citep{bastian21}. 
Such an encounter should have been able to provide the necessary gas reservoir for the formation of the young 
populations observed in these systems \citep{mckenzie18}.  

The latter scenario is expected to produce two discrete and well separated SF episodes. 
Hence, to shed new light on the physical mechanisms driving the formation of Liller~1 and Terzan~5, 
in this paper we present the first detailed analysis of the SFH of Liller~1.
To this aim, we have used deep optical HST and IR GeMS/GSAOI data used in combination 
with a synthetic CMD analysis to derive the SFH of the system.
The best-fit solution suggests that Liller~1 has been actively forming stars almost for its entire lifetime 
and we can identify three SF episodes.
The main episode started $12-13$ Gyr ago with a tail extending for up to $\sim3$ Gyr. 
This SF event is responsible for $\sim70\%$ of the present-day total mass of Liller~1.
The second peak occurred between 6 and 9 Gyr ago contributing to an additional $\sim15\%$ of the system's mass.
The most recent event started some 3 Gyr ago and stopped $\sim1$ Gyr ago, when a quiescent phase started.
Our analysis shows that the young population contributes at least to $\sim10\%$ of the total mass of Liller~1.
We also find that the best match with the observations is obtained by assuming a global average
metallicity [M/H]$=-0.5$ for the older stars, in quite good agreement with the few spectroscopic 
abundance measurements available in the literature \citep{origlia02}, 
and with [M/H]$\sim+0.2$ for stars that were born in the last $\sim3$ Gyr.

If we take the results of our analysis at a face value, 
we should conclude that the overall SFH complexity, 
the long time over which Liller~1 has been forming stars and the number of SF episodes suggest that Liller~1 
unlikely formed through the merger between an old globular cluster and a Giant Molecular Cloud, as recently 
proposed by \citet{bastian21}.
On the contrary, our findings provide further support to the idea that Liller~1 
is the surviving relic of a much more massive primordial structure, the so-called bulge fossil fragments, 
that contributed to the Galactic bulge formation. 
In this respect, it is worth stressing that stellar systems with original stellar masses 
larger than several $10^6 M_{\odot}$ are expected (see \citealt{bailin09}) to be able to retain a fraction of their SN ejecta even without the additional contribution of a dark matter halo, thus providing further support to the self-enrichment scenario.
Counter examples based on unresolved stellar populations as suggested by \citet{bastian21} can be biased by the uncertanties arising from the derivation of masses and stellar population properties from integrated quantities. Moreover, it is also questionable that the sole analysis of the integrated properties of Terzan~5 and Liller~1 could reveal their complexity and true nature.
\citet{bastian21} discuss the caveat that massive systems (such as Liller~1 and Terzan~5) in the dense inner Galaxy 
may be able to capture field stars characterized by a range of metallicities and ages. 
However, the expected number of 
captured stars is still small (of the order of a few percent), while the extended SF of Liller~1 
contributes to more than $30\%$ of its present-day mass.  
We should also note that the age spread needed to best-fit the old stellar population 
is not compatible with the assumption that it was originally a genuine GC. On the same line, the extended duration 
of the young SF event ($\Delta t\sim 2-3$ Gyr) appears to be hardly compatible 
with the encounter between a GMC and a massive GC.
    
We performed an ad-hoc simulation including `by hand' only two SF episodes at $t\sim12$ Gyr and $t<3$ Gyr.
Although the CMD derived by using this approach is still broadly compatible with the data, 
we find that the specific number count distribution of the RC and young MS
in the CMD (Figures~\ref{RCs} and \ref{lf}) cannot be properly reproduced with such a configuration,
but they are better fit by adopting a larger number of SF events lasting for a longer time. 

Our group is now conducting an extensive observational effort to obtain high-resolution spectroscopy of Liller~1
to probe the chemical abundance patterns of this system on a statistically significant sample possibly including 
the young population. 
This will certainly provide critical information to move the SFH analysis of this system to another step of detail.
On the same line, we are planning to apply the same analysis to the case of Terzan~5 and then compare the results
between the two systems.

\section*{Acknowledgements}
The research was funded from the project {\it Light-on-Dark} granted by MIUR through PRIN2017- 2017K7REXT.
E.V. acknowledges the Excellence Cluster ORIGINS Funded by the Deutsche Forschungsgemeinschaft
(DFG, German Research Foundation) under Germany's Excellence Strategy \-EXC\-2094\-39078331.
This work has made use of data from the European Space Agency (ESA) mission Gaia (https://www.cosmos.esa.int/gaia), 
processed by the Gaia Data Processing and Analysis Consortium (DPAC, https:// www.cosmos.esa.int/web/gaia/dpac/consortium). 
Funding for the DPAC has been provided by national institutions, in particular 
the institutions participating in the Gaia Multilateral Agreement.

\end{document}